\documentclass[doublecol]{epl2} 

\title{Spontaneous transition to a fast 3D turbulent reconnection regime}
\shorttitle{Spontaneous transition to a fast 3D turbulent reconnection regime} 

\author{Giovanni Lapenta\inst{1} \and Lapo Bettarini\inst{2}}
\shortauthor{Lapenta and Bettarini}

\institute{                    
  \inst{1} Centrum voor Plasma-Astrofysica, Departement Wiskunde, Katholieke Universiteit Leuven,Celestijnenlaan 200B, 3001 Leuven, Belgium\\
  \inst{2} Royal Observatory of Belgium, Ringlaan 3, 1180 Brussels, Belgium
}

\abstract{
We show how the conversion of magnetic field energy via magnetic reconnection can progress in a fully three-dimensional, fast, volume-filling regime. An initial configuration representative of many laboratory, space and astrophysical plasmas spontaneously evolves from the well-known regime of slow, resistive reconnection to a new regime that allows to explain the rates of energy transfer observed in jets emitted from accretion disks, in stellar/solar flare processes as well as in laboratory plasmas. This process does not require any pre-existing turbulence seed which often is not observed in the host systems prior to the onset of the energy conversion. The dynamics critically depends on the interplay of perturbations developing along the magnetic field lines and across them, a process possible only in three-dimensions. The simulations presented here are the first able to show this transition in a fully three-dimensional configuration.}

\begin{document}

\maketitle

\section{Introduction}

A continuous and volume-filling process converting magnetic energy into high energy particles is manifestly present in jets emitted from accretion disks around forming stars or black holes. In fact the synchrotron emission from the whole jet means that an active process is actually sustaining and reaccelerating a population of relativistic particles that hence emit. In absence of such a mechanism the relativistic population would be confined close to the energy source (e.g., the accretion disk) and quickly lose its energy by radiation as it moves away. The sustained production and acceleration of extremely energetic particles in vast macroscopic regions is also observed in solar system plasmas (Sun and Earth environments) and in laboratory. The mechanism underlying the above mentioned phenomena is yet to be fully understood, but it must clearly include two critical features: the acceleration process must be volume-filling  and it has to be sufficiently fast that the energy input from this mechanism matches the observed rates of energy conversion in Nature. 

Understanding the process of fast magnetic energy conversion remains a key task in plasma physics. Magnetic reconnection is the leading process long suggested to accomplish this task. In magnetic reconnection, field lines break and change their topological connection in a way that makes a piece of a field line connect with a piece of another field line~\cite{1,2}. In the process the magnetic energy is released in the form of kinetic energy, that is acceleration and heating. However, under any understood mechanism, on macroscopic scales, reconnection requires a dissipation process such as resistivity allowing the conversion, but in all observed natural and in most man-made conditions the level of dissipation measured or induced is several orders of magnitude too low. Also so-called anomalous dissipation processes have been traditionally invoked. However,  the microscopic mechanisms leading to the formation of the invoked anomalous effects depend on several factors and have not yet been conclusively identified. For example within the solar corona, the free non-potential energy of the (local) solar magnetic field is released via magnetic reconnection both during sudden and violent eruptive processes such as flares and coronal mass ejections (CME) or it can occur in a slowly quasi-steady way which may contribute to the coronal heating. Yet measured high temperatures of current-sheets formed in the aftermath of CME's~\cite{ 3, 4} and inferred characteristic dynamic (Alfv\'en) time scales throughout the different phases of solar explosive phenomena have pointed out a number of discrepancies between the level of plasma resistivity required in the solar corona, e.g. in the case of stationary current-sheets~\cite{5} wherein it is about $10^{11}-10^{12} \mbox{ m}^2 \, \mbox{s}^{-1}$, and the one deduced from theories of classical and anomalous resistivity, that is about $1 \mbox{ m}^2 \, \mbox{s}^{-1}$ and $10^{6}-10^{7} \mbox{ m}^2 \, \mbox{s}^{-1}$, respectively. This implies that a very efficient diffusion mechanism is at work and that neither the classical resistivity nor anomalous effects are necessarily governing this process. 

At the kinetic level, which is measured quantitatively as the radius of gyration of an ion particle in the ambient magnetic field, reconnection can be locally fast  without requiring the dissipation mechanisms needed on macroscopic scales. These scales have been recently analyzed especially for the jet originating from the relatively distant radio galaxy 3C303 ($z = 0.141$)~\cite{6}. The jet presents a knotty structure, that is brightly radiating spots alternate with dim segments, observed in both radio and X-ray bands and these knots may be considered as actual plasma features created near the central source and propagating along the jet axis. In this case the mechanism at work could be the reconnection of magnetic field lines~\cite{7}. It is conceivable that due to some yet unidentified mechanism kinetic-scale reconnection might be allowed to fill the whole jet and provide energy conversion at the macroscopic scales needed. However, very recent works have suggested that it is not necessary to invoke such extreme extrapolation from the observed scales of energy conversion to the minute scales where kinetic reconnection becomes active. We show here the first demonstration via numerical simulations of the ability of macroscopic processes of magnetic reconnection to achieve the two required goals needed to explain the observations:  the process is fast enough and it is volume filling. We remark that both results have long eluded the research. Fast reconnection within macroscopic models described by the viscous-resistive magnetohydrodynamics (MHD), which does not include kinetic processes or anomalous effects, have long been believed impossible. The consensus was that under these conditions reconnection would progress under the so-called Sweet-Parker (SP) rate and hence determine a very slow reconnection. This point of view changed in the last couple of years due to the discovery of a new regime where reconnection is unsteady and self-driven~\cite{8, 9, 10}. The typical state involved in the SP mechanism is destabilized and a new turbulent layer forms in its place. Yet due to the limitation of being two dimensional these discoveries still failed to explain the volume-filling feature of reconnection and according to them the fast reconnection layer remained localized. 

A possible way to overcome this last step was indicated by Lazarian and Vishniak who proposed that a fully three dimensional reconnection layer where turbulence is included could provide a macroscopic and fast reconnection scenario~\cite{11}. This idea had also been proposed in a cartoon-based speculative scenario by Shibata: the current-sheet undergoes recurrent secondary tearing instabilities~\cite{12, 13, 14, 15} occurring at different spatial scales and this leads to a fractal (i.e., scale-free) current-sheet structure, where many magnetic islands of different sizes (subject also to coalesce with each other) connect macroscopic and microscopic scales and fill the scale gap mentioned above~\cite{16}. The main idea is that the energy is supplied at the largest spatial scales and it is being redistributed via MHD turbulent cascade to smaller scales and finally dissipated. The first attempts to study this process were done in two dimensional simulations on the effect of MHD turbulence on magnetic-reconnection~\cite{17} and nowadays not only it is suggested to be a fundamental mechanism to model solar explosive phenomena~\cite{18}, but also recently it has been claimed to have obtained the first in situ evidences of magnetic reconnection in turbulent plasmas within the heliospheric environment~\cite{19}.

In the present paper we demonstrate quantitatively that fast, three dimensional and volume filling reconnection can be achieved spontaneously within the framework of the pure viscous-resistive MHD and even without invoking any externally driven turbulence. In fact the simulations presented here have an unprecedented low level of numerical diffusion that in previous studies eliminated the transition to the fast turbulent regime (replacing it with a fast laminar regime purely driven by numerical effects). Indeed a system completely laminar in conditions similar to those found in many astrophysical and laboratory systems is observed to settle at first in the well know slow SP regime, but its subsequent evolution spontaneously brings it through three stages. First, the destabilization of the SP regime previously studied in two dimensions repeats itself and even in three dimensionsional simulations the system still maintains a two-dimensional behavior. Secondly, an explosive growth of instabilities in the third dimension breaks the translational invariance and makes the evolution fully three-dimensional. In the third stage, a new regime is eventually observed, where the reconnected states are fully mixed and the process encompasses the whole domain. The transition in 3D, unlike in 2D, allows to reach regions of macroscopic size and providing the needed macroscopic channel to explain the observed amounts of energy conversion. Furthermore the dynamics critically depends on the interplay of perturbations developing along the magnetic field lines and across them, a process possible only in three-dimensions. Both the volume-filling and the fast-progressing features of magnetic reconnection are within the same model and without requiring extra physical processes such as anomalous resistivity.

\section{Method}

In order to reach the needed high resolution and to reduce the influence of spurious numerical effects, we use a code with a fully-implicit (in time) particle (or FLIP) algorithm to solve an Eulerian-Lagrangian formulation of three dimensional resistive and viscous MHD equations, FLIP3D-MHD~\cite{20}. This algorithm solves the MHD equations in four steps and it uses both a particle and a grid description of the fluid: 
\begin{enumerate}
\item particle data are interpolated to a Lagrangian grid co-moving with the fluid;
\item Lagrangian finite difference equations are solved on the moving grid to advance the fields;
\item changes in the grid data over a time step are interpolated to the particles to advance the particle data;
\item a new grid is generated (to avoid the tangling of a Lagrangian mesh) and the particles are interpolated to the grid for the next time step 
\end{enumerate}
The most important difference between FLIP and earlier particle-in-cell methods~\cite{21} is in step 3. By interpolating changes in the solution from the grid to the particles, rather than by replacing the particle data with the new grid solution. Consequently, computational diffusion decreases with the time step, $\Delta t$. This change guarantees consistency, even with higher order interpolation, and reduces computational diffusion below that achievable with Eulerian finite difference methods with a similar range of applicability. 

The particle-in-cell (PIC) method is Galilean invariant~\cite{22} and FLIP extends this property to the MHD equations. Galilean invariance is achieved by making the magnetic field a particle variable. Changes in the magnetization are interpolated to the particles to calculate the new particle magnetic moment. This last step introduces some computational diffusion. However, the diffusion is small compared with that produced by, e.g., van Leer convection~\cite{20}. Spatial differencing insures that differencing at grid vertices and cell centers are conjugate operations~\cite{23}. Thus momentum is conserved, and energy is conserved with implicit differencing in time.

FLIP has been validated by comparisons of analytically solvable problems, such as the Rayleigh-Taylor instability, confined eddy, and Kelvin-Helmholtz instability~\cite{20, 24, 25}, and by comparisons with other numerical computations. 

In the simulations presented here, the visco-resistive MHD equations have a uniform viscosity and resistivity parametrized by the global Lundquist number, $S$, and kinetic Reynolds number, $R_M$, both set to $10^4$. We define a stream-wise direction ($X$) with periodic boundary conditions at $X$ = 0 and $L_X$, a cross-stream ($Z$) and span-wise ($Y$) direction with, respectively,  reflecting and periodic boundary conditions at $Z (Y) = 0$ and $Z (Y) = L_Z (L_Y)$. We consider $360$ (in $X$) x $60$ (in $Y$) x $240$ (in $Z$) Lagrangian markers arrayed initially in a $3$ x $3$ x $3$ uniform formation in each of the $120$ x $20$ x $80$ cells of our numerical grid. The mathematical formulation of the problem include an ideal equation of state with polytropic index $\gamma$ such that $p = \rho (\gamma - 1) I$ where $p$, $\rho$ and $I$ are respectively the kinetic pressure, the density and the enthalpy of the plasma. The basic formula has the main field component reversed, $B_X (Z) = B_0 \tanh(Z/L)$, over a characteristic length scale $L$. Additional component can be considered to obtain the so-called guide field configuration (with a uniform magnetic field component in the direction orthogonal to the gradients and the main field), $B_Y (Z) = \alpha B_0$ ($\alpha = \mbox{constant}$), or the force-free configuration characteristic of situations where the plasma is at its minimum energy state and all forces are absent, $B_Y (Z)= B_0 /\cosh(Z/L)$. While all cases were considered in the present study, the results will focus on the simples case of $B_Y = 0$.

\section{Sweet-Parker layer and its destabilization: magnetic connectivity and energy evolution}

When regions of different magnetic field orientation come in contact, the simplest case being two regions of opposite polarity,  the possibility arises for the  process of reconnection. For astrophysical jets, this could be the case of the field generated from the accretion disk and carried along the jet by the Poynting flux. Magnetic field lines coming from the disk must return to it (assuming the classical absence of magnetic monopoles), leading to regions of different magnetic field orientation. Moreover the rotation of the disk transmits a rotation also to the field lines, which are hence wound around the jet, and this facilitates the insurgence of regions of reconnection~\cite{26}. Similar situations are observed in the solar system as well, i.e. regions of opposing polarity coming in contact in the photosphere of the Sun, opposing magnetic field lines in the solar wind or in the magnetosphere (during periods of southward directed interplanetary magnetic field compared with the northward directed Earth field), or the Earth field lines bent and push together in the Earth magnetic tail. In laboratory, such configurations are also present by design in fusion devices and in devices especially designed to study the process of magnetic reconnection~\cite{27}.

\begin{figure}
\centering\includegraphics[width=0.8\columnwidth]{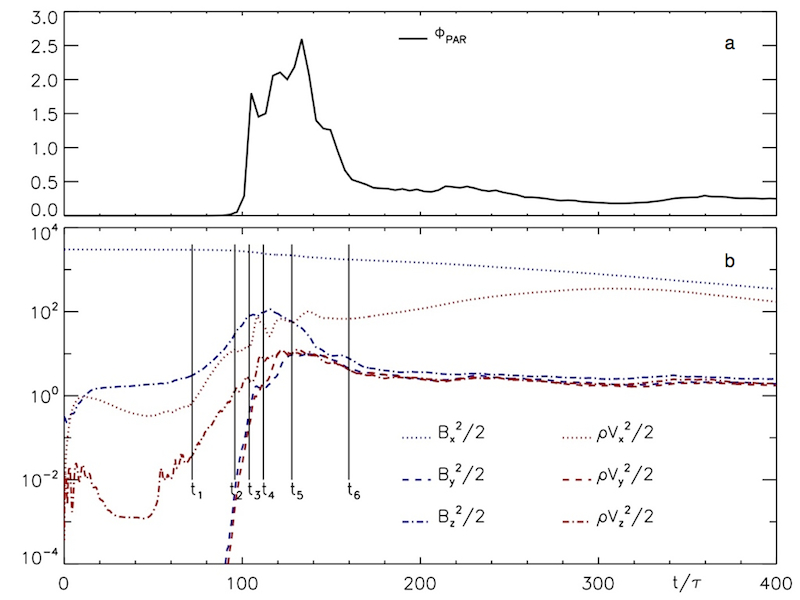}
\caption{\textbf{General evolution of the current-sheet dynamics: electric field and different contributions to the energy during the simulation performed with FLIP3D-MHD}. Panel (a) shows the parallel electric field, $\Phi_{PAR} = \mathbf{E} \cdot \mathbf{B}/|\mathbf{B}|$. Panel (b) shows the different magnetic and kinetic contributions to the system energy during current-sheet reconnection dynamics. Moreover the system is sampled at six remarkable instants indicated by $t_1 = 72$, $t_2 = 96$, the time interval between $t_3 = 104$ and $t_4 = 112$, $t_5 = 128$, and $t = 160$ clearly shown in the figure.}
\label{figure1}
\end{figure}
In the panel (a) of  Fig.~\ref{figure1} the parallel electric field ($= \Phi_{PAR} = \mathbf{E} \cdot \mathbf{B}/|\mathbf{B}|$) integrated in the whole numerical domain is shown as a function of time: the rate of changes in the magnetic connectivity is associated with the maximum value of $\Phi_{PAR}$~\cite{28}. At $t \approx 72$ (dimensionless units of time measured with respect to the typical Alfv\'en time) the system experiences an abrupt switch from an initial slow and laminar reconnection regime to a fast reconnection process. The parallel electric field evolution remarkably underlines the spontaneous transition to the fast regime lasting from $t = 72$ to about $t = 96$ where the (first) maximum of $\Phi_{PAR}$ is reached. The second, and highest, peak describes a third reconnection phase occurring now along the span-wise direction ($Y$), that is the direction orthogonal to the first tearing plane: it can be present only in fully 3D configurations and, as we show, it corresponds to a macroscopic turbulent configuration. In fact this last phase dispalys a volume-filling kinking instability that drives the system evolution towards a final chaotic state. In the panel (b) the evolution of the magnetic and kinetic contributions to the system energy is considered. The evolution is divided into several time intervals from $t_1 = 72$ to $t_6 = 160$ (dimensionless units of time) useful in the analysis of the subsequent data. The initial reservoir of magnetic energy, which corresponds essentially to the $B^2_X/2$ contribution, is rapidly converted to kinetic energy.

\begin{figure}
\centering\includegraphics[width=0.8\columnwidth]{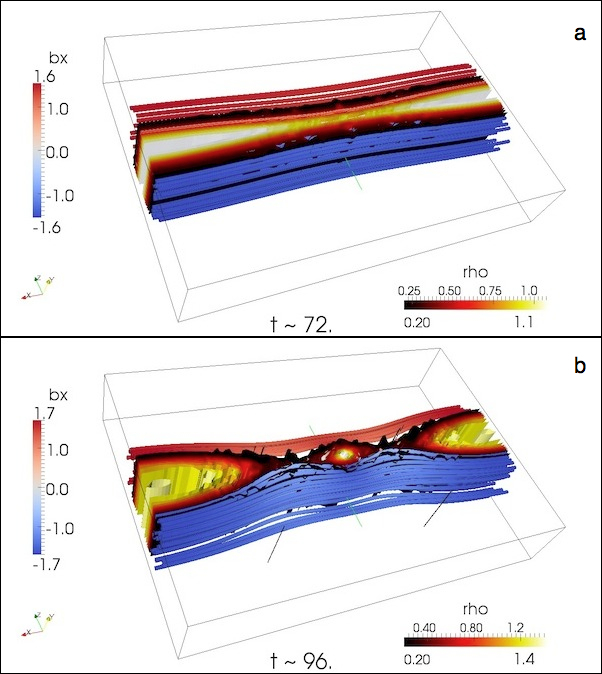}
\caption{\textbf{Volume iso-surface of the density in the region of the current-sheet at the centre of our numerical domain and magnetic field lines}. The blue-red color table for the magnetic field lines describes the intensity of the $X$-component. The panel (a) shows the system at $t \approx 72 \, (t_1)$ when the disruption in multiple pieces of the current-sheet starts. The panel (b) shows the system at $t \approx 96 \, (t_2)$ when coalescence process leads to the formation of an unique plasmoid at the centre of the box.}
\label{figure2}
\end{figure}
\begin{figure}
\centering\includegraphics[width=0.8\columnwidth]{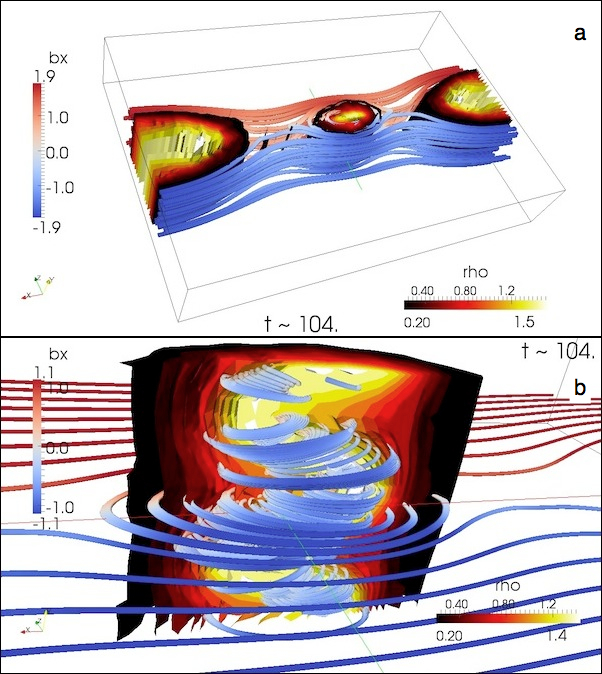}
\caption{\textbf{Volume iso-surface of the density in the region of the current-sheet at the centre of our numerical domain and magnetic field lines}. The blue-red color table for the magnetic field lines describes the intensity of the $X$-component. The panel (a) shows the system at $t \approx 104 \, (t_3)$ and in the panel (b) a detailed clip of the central magnetically-confined higher-density region wherein It is evident the sinuous mode of the kinking dynamics of the structure and the complex closed loops of the magnetic field following the instability dynamics of the island.}
\label{figure3}
\end{figure}
\begin{figure}
\centering\includegraphics[width=0.8\columnwidth]{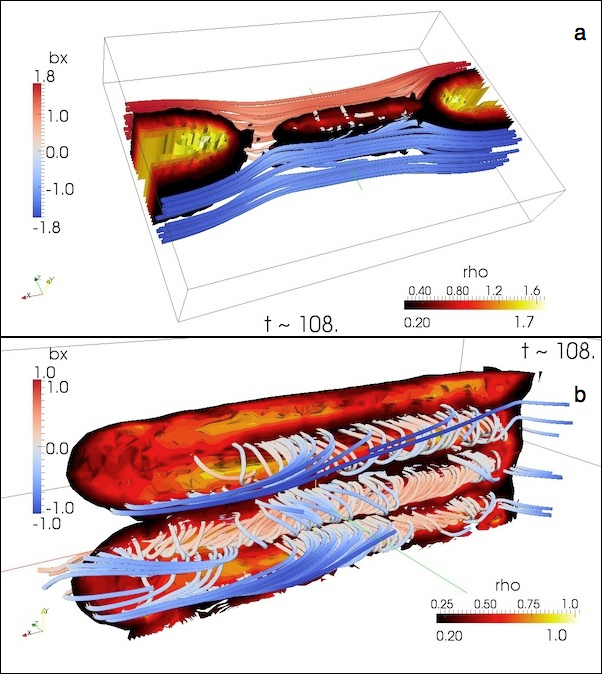}
\caption{\textbf{Volume iso-surface of the density in the region of the current-sheet at the centre of our numerical domain and magnetic field lines }. The blue-red color table for the magnetic field lines describes the intensity of the $X$-component. The panel (a) shows the system at $t \approx 108$, whereas a detailed clip of the central island is shown in the panel (b). It is evident the final evolution (of the dominating mode) of the kinking instability driving the system and the complex configuration assumed by the magnetic field that is anchored at the structured.}
\label{figure4}
\end{figure}
In Fig.~\ref{figure2}-\ref{figure4} the volume iso-surface of the density together with the magnetic field lines (colored according to the value of the $X$-component of the magnetic field) are shown for $t \approx 72 \, (t_1)$ and $t = 96 \, (t_2)$ (see Fig.~\ref{figure1} for the definition of $t_i$ with $i = 1,...,6$). As shown in Fig.~\ref{figure2}, panel (a), at $t = t_1$, the current-sheet, thus far undergoing a slow Sweet-Parker (laminar) reconnection, destabilizes and tears into multiple-islands separated by  smaller-scale current-sheets: the system chaotically evolves towards smaller and smaller length-scales, but it still maintains a two-dimensional configuration on any $X$-$Z$ plane of the numerical domain. Previous studies have analyzed this phase, its fractality, turbulence and its scaling~\cite{9, 29, 30}. The panel (b) of the same figure shows that at $t = t_2$, the coalescence instability among the secondary islands determines an inverse cascade till a three secondary-island system is reached. So far, the system still preserves its two-dimensional behavior.

\begin{figure}
\centering\includegraphics[width=0.8\columnwidth]{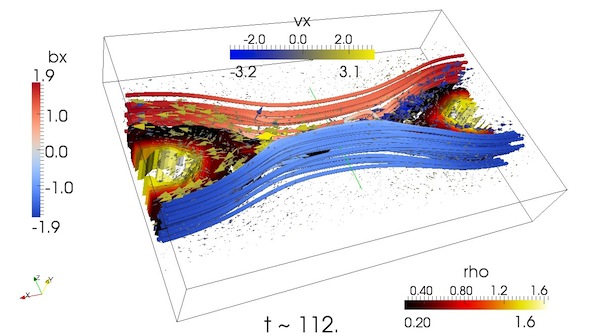}
\caption{\textbf{Volume iso-surface of the density in the region of the current-sheet at the centre of our numerical domain, magnetic field lines, and velocity field}. The system is considered at $t \approx 112 \, (t_4)$, panel (a). The color tables for the magnetic field lines and the velocity arrow glyphs describe the intensity of the x-component. The disruption of the central island produces the expulsion of higher-density regions transported by the increased velocity field.}
\label{figure5}
\end{figure}

At this point, however, the system rapidly undergoes  a full three-dimensional instability as shown in both panels of Fig.~\ref{figure3} referring to the instant $t \approx 104 \, (t_3)$. In the panel (a), the whole box is considered, while on the right a blow up is shown of the central region. The density structure is strongly kinked  while the overall macroscopic configuration of the magnetic field topology is largely preserved to its previous invariance along the $Y$ direction. Several secondary instability modes gain ground onto the previously two-dimensional structure and drive a full space destabilization. The magnetic field loops around the high density regions following the plasma dynamics. The instability proceeds fast and in a few (dimensionless) crossing times of the current layer the central island is completely distorted by the kinking mode as shown in Fig.~\ref{figure4}, panel (a). In the panel (b) of the same figure, a detailed view is presented of the central part of our numerical domain where the system is rearranged along the span-wise direction ($Y$) with the magnetic field completely tangled to the plasma in a configuration highly unstable again to magnetic reconnection. Indeed the system tears but in the orthogonal direction with respect the initial settings: reconnection spreads volume-wide. The system breaks and the disruption of the central island determines the expulsion of higher-density regions transported by an increased velocity field as shown in Fig.~\ref{figure5} for $t \approx 128 \, (t_4)$. This fragmentation accelerates along the overall magnetic field lines and hits the two islands located at the side of the numerical domain.

\section{Mixing of the reconnected fluxes and fully turbulent state}

\begin{figure}
\centering\includegraphics[width=\columnwidth]{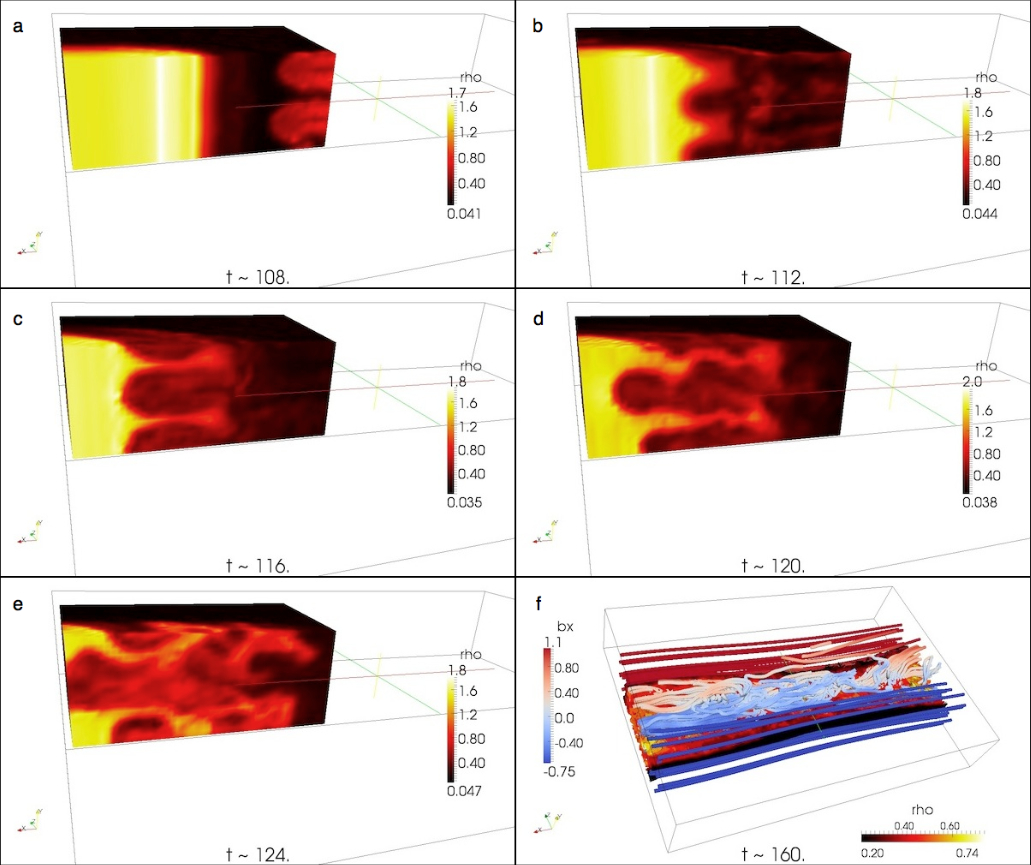}
\caption{\textbf{Clip of the volume iso-surface of the density of (one of the two) higher-density regions at the side of our domain and final evolution of the magnetic field \textit{plus} density}. In the panel from (a) to (e) we can observe that the instability dynamics is triggered probably by the coalescence with the fragments expelled by the central island under the action of the increased $X$ component of the velocity. The Panel (f) show the final chaotic evolution of the system at the end of the simulation (the blue-red color table for the magnetic field lines describes the intensity of the $X$-component).}
\label{figure6}
\end{figure}

The turbulence and speed of the process of reconnection leads to an additional consequence: the destabilization of the downstream regions to a further Rayleigh-Taylor (RT) instability. As the plasma from the reconnection region is expelled, it collects downstream forming two regions of higher density, called plasmoids. As reconnection jets slow down and form the plasmoid, the acceleration field points contrary to the jet, while the jet pushes towards the plasma that is piling up downstream. A configuration with acceleration and density gradient with opposite direction (the density increases along the reconnection jet, the acceleration points in the opposite direction) is unstable to the RT instability. The turbulence of the reconnection processes feeds this instability by providing the needed perturbations to allow the mode to grow. Conversely, the RT profoundly disturbs the 2D symmetry facilitating the mixing in the 3D domain. Figure~\ref{figure6}, panel (a)-(f), shows the characteristic fingers produced by the attempt of the heavy fluid to fall into the direction of the acceleration, just as a heavy fluid would do when layered over a light fluid in a gravity field. The evolution of the instability  completely destroys the overall structure of the system leading to a completely chaotic configuration as shown in ~Fig.~\ref{figure6}, panel (f), where the density contour and magnetic field lines are shown at $t \approx 160 \, (t_6)$.

\acknowledgments
The present work is supported by the Onderzoekfonds KU Leuven (Research Fund KU Leuven) and by  the European Commission under the Seventh Framework Programme (FP7/2007-2013) under the grant agreement no. 218816 (SOTERIA, www.soteria-space.eu).

\end{document}